\documentclass[prd,twocolumn,showkeys,amsmath,amssymb]{revtex4-1}
\usepackage{graphicx}
\usepackage{epstopdf}
\usepackage{multirow}
\usepackage{soul}
\usepackage{dcolumn}
\usepackage{bm}
\usepackage{mathrsfs}
\usepackage[colorlinks,citecolor=blue,anchorcolor=red,menucolor=red,linkcolor=red,filecolor=red,runcolor=red,urlcolor=blue,frenchlinks=red]{hyperref}
\usepackage{color}
\usepackage{array}
\usepackage{subfig}
\usepackage{float}
\usepackage{overpic}
\usepackage{caption}

\allowdisplaybreaks[3]

\newcommand{\mev}{\textrm{ MeV}}

\begin{document}

\title{Hadronic molecules with four charm or beauty quarks}

\author{Wen-Ying Liu}
\author{Hua-Xing Chen}
\email{hxchen@seu.edu.cn}
\affiliation{School of Physics, Southeast University, Nanjing 210094, China}

\begin{abstract}
We apply the extended local hidden gauge formalism to study the meson-meson interactions with the quark constituents $cc\bar c\bar c$, $cc\bar c\bar b/\bar c\bar c cb$, $cc\bar b\bar b/\bar c\bar c bb$, $bb\bar c\bar b/\bar b\bar b cb$, and $bb\bar b\bar b$, where the exchanged mesons are the fully-heavy vector mesons $J/\psi$, $B_c^*$, and $\Upsilon$. We solve the coupled-channel Bethe-Salpeter equation to derive two poles in the $bb\bar{c}\bar{b}$ system and two poles in the $cc\bar{c}\bar{b}$ system. There are also four charge-conjugated poles in the $\bar{b}\bar{b}cb$ and $\bar{c}\bar{c}cb$ systems. In the $bb\bar{c}\bar{b}$ system, one pole corresponds to a sub-threshold bound state when the cutoff momentum is set to $\Lambda > 850 \mev$. The other pole in this system corresponds to a sub-threshold bound state when $\Lambda > 1100 \mev$. In the $cc\bar{c}\bar{b}$ system, the two poles correspond to sub-threshold bound states only when $\Lambda > 1550 \mev$ and $\Lambda > 2650 \mev$. This makes them difficult to identify as deeply-bound hadronic molecules. We propose investigating the two poles of the $bb\bar{c}\bar{b}$ system in the $ \mu^+\mu^-B_c^-$channel at the LHC.
\end{abstract}


\maketitle

\section{Introduction}
\label{intro}

In recent decades, the study of exotic hadrons has gradually become a focal point in hadron physics. Some exotic hadrons exhibit multiquark compositions, such as the compact tetraquark states and the meson-meson molecular states~\cite{Liu:2019zoy,Chen:2022asf,Guo:2017jvc,Brambilla:2019esw,Esposito:2016noz,Lebed:2016hpi,Ali:2017jda}. The picture of hadronic molecules has achieved significant success in the light-quark sector~\cite{Oller:1997ti,Oller:1998hw,Oset:1997it,Jido:2003cb,Bruns:2010sv,Garcia-Recio:2003ejq,Hyodo:2002pk}, which can be used to explain many resonances, such as the $f_0(980)$ and $a_0(980)$, etc. Besides, many hidden-charm pentaquark states observed in the past decade can be interpreted as the hadronic molecules that are dynamically generated through the meson-baryon interactions within the local hidden gauge framework~\cite{Wu:2010jy,Wu:2010vk,Chen:2015sxa,He:2015cea,Xiao:2013yca,Roca:2015dva,Liu:2015fea,Uchino:2015uha}. In recent years, several exotic structures in the di-$J/\psi$ invariant mass spectrum have been reported by the LHCb, CMS, and ATLAS collaborations~\cite{LHCb:2020bwg,ATLAS:2023bft,CMS:2023owd}, including the $X(6200)$, $X(6600)$, $X(6900)$, and $X(7200)$. These structures are good candidates for the fully-charmed tetraquark states. Extensive theoretical investigations have been performed to elucidate their nature~\cite{liu:2020eha,Tiwari:2021tmz,Lu:2020cns,Faustov:2020qfm,Zhang:2020xtb,Li:2021ygk,Bedolla:2019zwg,Weng:2020jao,Liu:2021rtn,Giron:2020wpx,Karliner:2020dta,Zhao:2020zjh,Mutuk:2021hmi,Wang:2021kfv,Wang:2020ols,Ke:2021iyh,Zhu:2020xni,Jin:2020jfc,Yang:2021hrb,Albuquerque:2020hio,Albuquerque:2021erv,Wu:2022qwd,Asadi:2021ids,Yang:2020wkh,Feng:2020riv,Ma:2020kwb,Maciula:2020wri,Goncalves:2021ytq,Wang:2020gmd,Esposito:2021ptx,Zhuang:2021pci,Zhao:2020nwy,Becchi:2020uvq,Sonnenschein:2020nwn,Zhu:2020snb,Wan:2020fsk,Gordillo:2020sgc,Liu:2020tqy,Majarshin:2021hex,Kuang:2022vdy,Wang:2021mma,Chen:2016jxd,Czarnecki:2017vco}, some of which attempted to explore their nature as molecular states~\cite{Guo:2020pvt,Cao:2020gul,Gong:2020bmg,Dong:2021lkh,Ortega:2023pmr,Wang:2022yes,Zhou:2022xpd}, but a definitive and conclusive understanding of their nature remains elusive.

Previous theoretical studies on the fully-heavy tetraquark states mainly focus on the interpretation of compact tetraquark states, while there are not so many studies based on the interpretation of hadronic molecular states. This is because the exchanged hadrons of these systems have quite large masses, such as the fully-heavy vector mesons $J/\psi$, $B_c^*$, and $\Upsilon$ exchanged in the $c b \bar c \bar b$ system within the extended local hidden gauge formalism, so their induced interactions are significantly suppressed. In Ref.~\cite{Liu:2023gla} we studied the $c b \bar c \bar b$ system to explore the existence of the fully-heavy hadronic molecules $B_c^{(*)}\bar B_c^{(*)}$. Within the extended local hidden gauge framework, we found that the two fully-heavy mesons $B_c^{(*)}$ and $\bar B_c^{(*)}$ are able to form a bound state by exchanging the relatively lighter meson $J/\psi$.

In this paper we apply the extended local hidden gauge formalism to further investigate the fully-heavy hadronic molecules existing in the $cc\bar c\bar c$, $cc\bar c\bar b/\bar c\bar c cb$, $cc\bar b\bar b/\bar c\bar c bb$, $bb\bar c\bar b/\bar b\bar b cb$, and $bb\bar b\bar b$ systems. By solving the coupled-channel Bethe-Salpeter equation, we evaluate the hadronic molecules generated from the meson-meson interactions in these systems. Our results indicate the possible existence of two bound states in the $bb\bar c\bar b$ system, along with two charge-conjugated states in the $\bar b\bar b cb$ system. However, their manifestation depends on the cutoff momentum, where they may appear as threshold effects. Both structures share the same spin-parity quantum number $J^P=1^+$, and can potentially be observed in the $\mu^+\mu^-B_c^-$ channel at LHC. Besides, we find two poles in the $cc\bar c\bar b$ system  (with two charge-conjugated poles in the $\bar c\bar c cb$ system), but it is difficult to identify them as deeply-bound hadronic molecules.

This paper is organized as follows. In Sec.~\ref{sec:formlism} we apply the local hidden gauge formalism to derive the potentials for the interactions of the $cc\bar c\bar c$, $cc\bar c\bar b/\bar c\bar c cb$, $cc\bar b\bar b/\bar c\bar c bb$, $bb\bar c\bar b/\bar b\bar b cb$, and $bb\bar b\bar b$ systems. Based on the obtained results, we solve the coupled-channel Bethe-Salpeter equation in Sec.~\ref{sec:result} to extract the poles, some of which may qualify as fully-heavy hadronic molecules. A brief summary is given in Sec.~\ref{sec:con}.

\section{formalism}
\label{sec:formlism}

\begin{figure}[H]
    \centering
    \includegraphics[width=0.95\linewidth]{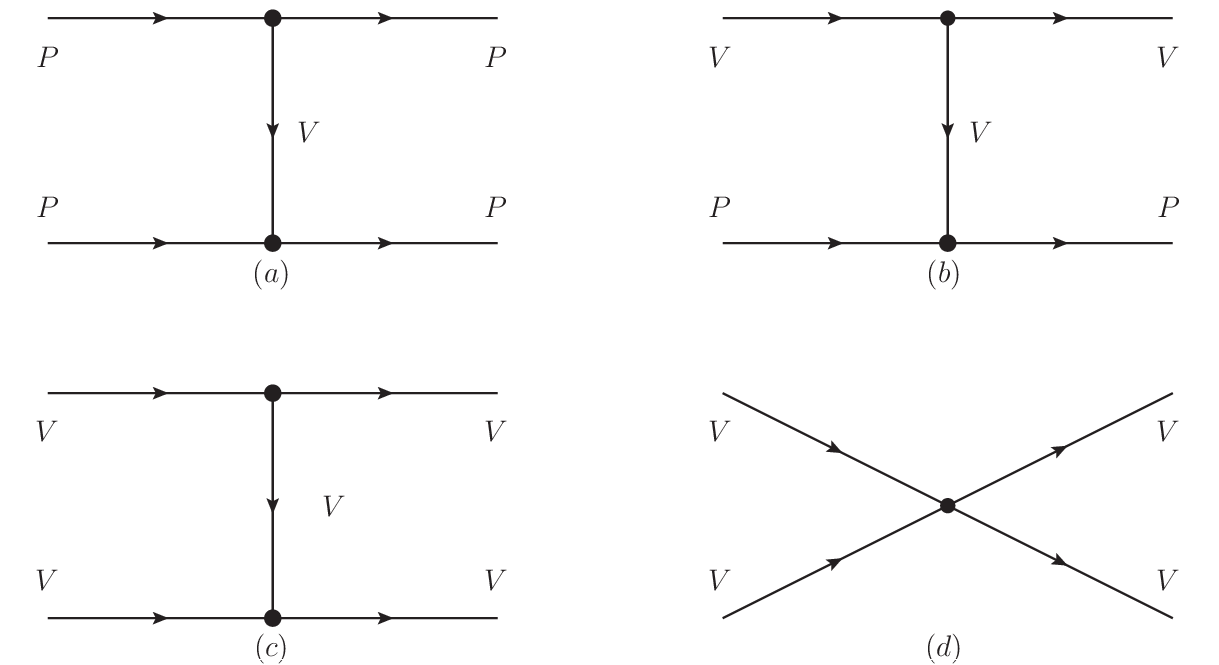}
    \caption{The meson-meson interactions arising from the vector meson exchange: (a) between two pseudoscalar mesons, (b) between one vector meson and one pseudoscalar meson, and (c) between two vector mesons. The subfigure (d) describes the contact term connecting four vector mesons.}
    \label{fig-lagrangians}
\end{figure}

In Ref.~\cite{Liu:2023gla} we have applied the extended local hidden gauge formalism to study the interactions of the $c b \bar c \bar b$ system. In this section we follow the same approach to study the $cc\bar c\bar c$, $cc\bar c\bar b/\bar c\bar c cb$, $cc\bar b\bar b/\bar c\bar c bb$, $bb\bar c\bar b/\bar b\bar b cb$, and $bb\bar b\bar b$ systems. Note that the validity of the extended local hidden gauge formalism in these systems is still questionable. Within the framework of local hidden gauge ~\cite{Bando:1987br,Meissner:1987ge}, the vector mesons are considered to act as gauge bosons, transmitting interactions. This mechanism successfully describes many low-energy interactions~\cite{Wu:2010jy,Oset:2010tof,Aceti:2014uea,Geng:2008gx,Nagahiro:2008cv}, especially in processes dominated by the exchange of light vector mesons. Unfortunately, it is not clear whether this mechanism remains effective in the heavy flavor region. An extension of this method has also been developed to study molecular state candidates containing heavy quarks~\cite{Wu:2010jy,Roca:2015dva,Molina:2009ct}. In these cases, generally, the exchange of light vector mesons and heavy vector mesons is allowed simultaneously. Taking the vector meson-vector meson interaction in the $c \bar c s \bar s$ system as an example, this system involves two coupling channels: $D_s^* \bar{D}_s^*$ and $J/\psi \phi$. In the $D_s^* \bar{D}_s^*$ channel, the exchange of $\phi$ and $J/\psi$ occurs simultaneously. Compared to the $\phi$ exchange, the $J/\psi$ exchange is considered to be secondary due to its larger mass. However, things are somewhat different in the $J/\psi \phi$ channel. The minimal constituents for the exchanged meson should be $c \bar{c} s \bar{s}$, which we do not consider in the local hidden gauge formalism. Such type of exchange mechanism should also be suppressed due to their involvement in the exchange of four quarks. On the other hand, $J/\psi \phi$ channel can couple to $D_s^* \bar{D}_s^*$ channel, which allows for the exchange of the heavy vector meson $D_s^*$ between these two channels. Although the exchange of heavy vector mesons should be relatively small, it plays a role in this scenario. Under this assumption, results consistent with experiments can be obtained~\cite{Molina:2009ct}. So the present study as well as Ref.~\cite{Liu:2023gla} serve as pioneer researches to investigate the fully-heavy hadronic molecules. Within this framework, the interactions are primarily contributed by the vector meson exchange, as depicted in Fig.~\ref{fig-lagrangians}. The corresponding Lagrangians are written as:
\begin{eqnarray}
\label{eq:lagrangians}
    \nonumber
    \mathcal{L}_{VPP} &=& -ig \, \langle[P,\partial_{\mu} P] V^{\mu}\rangle \, ,
    \\[1.5mm]
    \mathcal{L}_{VVV} &=& ig \, \langle(V^{\mu}\partial_{\nu}V_{\mu}-\partial_{\nu}V^{\mu}V_{\mu})V^{\nu}\rangle \, ,
    \label{eq-lagrangians}
    \\
    \nonumber
    \mathcal{L}_{VVVV} &=& \frac{g^2}{2}\langle V_{\mu}V_{\nu}V^{\mu}V^{\nu}-V_{\nu}V_{\mu}V^{\mu}V^{\nu}\rangle \, ,
\end{eqnarray}
where
\begin{equation}
    P =
    \left(\begin{array}{cc}
            \eta_c & B_c^+  \\
            B_c^-  & \eta_b
        \end{array} \right) , ~~~
    V =
    \left( \begin{array}{cc}
            J/\psi       & B_c^{*+} \\
            B_c^{*-}     & \Upsilon
        \end{array} \right) \, .
\end{equation}
The coupling constant $g$ is generally defined as $g={M_V}/({2f_P})$, with $M_V$ the mass of the exchanged vector meson and $f_P$ the decay constant of its corresponding pseudoscalar meson. Since the charm and bottom quarks do not form a flavor $SU(2)$ symmetry, we can not use an overall parameter. For the exchange of the $J/\psi$, $B_c^*$, and $\Upsilon$ mesons, we respectively use
\begin{eqnarray}
\nonumber
& M_{J/\psi} = 3096.9 \mbox{ MeV~\cite{pdg},   } f_{\eta_c}=387/\sqrt{2} \mbox{ MeV~\cite{Becirevic:2013bsa}}, &
\\
&  M_{B_c^*} = 6331 \mbox{ MeV~\cite{Mathur:2018epb},   } f_{B_c}=427/\sqrt{2} \mbox{ MeV~\cite{McNeile:2012qf}}, &
\\ \nonumber
& M_{\Upsilon} = 9460.4 \mbox{ MeV~\cite{pdg},   } f_{\eta_b}=667/\sqrt{2} \mbox{ MeV~\cite{McNeile:2012qf}}. &
\end{eqnarray}
Besides, we use $g^4 = g_{V_1}g_{V_2}g_{V_3}g_{V_4}$ for the contact term, with $V_{1\cdots4}$ the four connected vector mesons.

We derive the following interaction potentials from the Lagrangians given in Eq.~(\ref{eq-lagrangians}) as,
\begin{eqnarray}
    \label{eq:vpp}
    V_{PP}(s) &=& C_{PP}^t \times g^2 (p_1+p_3) (p_2+p_4)
    \\ \nonumber &+& C_{PP}^u \times g^2 (p_1+p_4) (p_2+p_3) \, ,
\\
    \label{eq:vvp}
    V_{VP}(s) &=& C_{VP}^t \times g^2 (p_1+p_3) (p_2+p_4) ~ \epsilon_1 \cdot \epsilon_3
    \\ \nonumber &+& C_{VP}^u \times g^2 (p_1+p_4) (p_2+p_3) ~ \epsilon_1 \cdot \epsilon_3 \, ,
\\
    \label{eq:vvv}
    V_{VV}(s)&=&V^{ex}_{VV}(s)+V^{co}_{VV}(s) \, ,
\\
    \label{eq:contact}
    V^{ex}_{VV}(s) &=& C_{VV}^t \times g^2 (p_1+p_3) (p_2+p_4)\epsilon_1 \cdot \epsilon_3 \epsilon_2 \cdot \epsilon_4
    \\ \nonumber &+& C_{VV}^u \times g^2 (p_1+p_3) (p_2+p_4)\epsilon_1 \cdot \epsilon_4 \epsilon_2 \cdot \epsilon_3 \, ,
\end{eqnarray}
where $p_1$ and $p_2$ are four-momenta of the incoming mesons; $p_3$ and $p_4$ are four-momenta of the outgoing mesons; $\epsilon_1$ and $\epsilon_2$ are polarization vectors of the incoming mesons; $\epsilon_3$ and $\epsilon_4$ are polarization vectors of the outgoing mesons. We use the subscripts $PP$, $VP$, and $VV$ to denote the pseudoscalar-pseudoscalar, vector-pseudoscalar, and vector-vector sectors, respectively. We use the superscripts $t$, $u$, and $co$ to denote the vector meson exchange in the $t$ and $u$ channels as well as the contact term, respectively.
 Note that the coefficient $C_{VP}^u$ is zero because we only consider the vector meson exchange in the present study. Actually, there also exists the pseudoscalar meson exchange. As detailedly evaluated in Ref.~\cite{Aceti:2014kja,Oset:2022xji}, the contribution of the pseudoscalar meson exchange is negligible near the threshold, compared to the vector meson exchange. This is because the amplitude of the vector meson exchange is proportional to the energy of the external meson, while the amplitude of the pseudoscalar meson exchange is proportional to the three-momentum of the external meson that is always negligible near the threshold.

We further derive the scattering amplitudes from Eqs.~(\ref{eq:vpp}-\ref{eq:contact}) by solving the Bethe-Salpeter equation as,
\begin{equation}
    T_{PP/VP/VV} = \left({\bf 1} - V_{PP/VP/VV} \bullet G\right)^{-1} \bullet V_{PP/VP/VV} ,
    \label{eq-bseq}
\end{equation}
where $G(s)$ is the diagonal loop function, whose expression for the $i$th channel is
\begin{equation}
    G_{ii}(s) = i\int{\frac{d^4q}{(2\pi)^4}\frac{1}{q^2-m_1^2+i\epsilon}\frac{1}{(p-q)^2-m_2^2+i\epsilon}} \, .
    \label{eq-G}
\end{equation}
In the above expression, $m_{1,2}$ are the masses of the two mesons involved in this channel, and $s = p^2$ with $p$ the total four-momentum. We apply the cutoff method to regularize it as
\begin{equation}
    G_{ii}(s) = \int_0^{\Lambda} \frac{d^3q}{(2\pi)^3} \frac{\omega_1 + \omega_2}{2\omega_1\omega_2} \frac{1}{s-(\omega_1+\omega_2)^2+i\epsilon} \, ,
    \label{eq-cutoff-G}
\end{equation}
where $\omega_1 = \sqrt{m_1^2+\vec{q}^{\,2}}$, $\omega_2 = \sqrt{m_2^2+\vec{q}^{\,2}}$, and $\Lambda$ is the cutoff momentum.

We shall calculate the coefficient matrices $C_{PP/VP/VV}^{t/u/co}$ in the following subsections, separately for the $cc\bar c\bar c$, $cc\bar c\bar b/\bar c\bar c cb$, $cc\bar b\bar b/\bar c\bar c bb$, $bb\bar c\bar b/\bar b\bar b cb$, and $bb\bar b\bar b$ systems. Before doing this, we summarize their relevant coupled channels in Table~\ref{tab:coupled-channels}.

\begin{table}[htb]
    \renewcommand{\arraystretch}{1.4}
    \centering
    \caption{Coupled channels considered for the $cc\bar c\bar c$, $cc\bar c\bar b/\bar c\bar c cb$, $cc\bar b\bar b/\bar c\bar c bb$, $bb\bar c\bar b/\bar b\bar b cb$, and $bb\bar b\bar b$ systems.}
    \setlength{\tabcolsep}{2mm}{
        \begin{tabular}{c|ccc}
            \hline\hline
            Constituent      & $PP$ sector    & $VP$ sector                       & $VV$ sector
            \\ \hline\hline
            $cc\bar c\bar c$ & $\eta_c\eta_c$ & $J/\psi\eta_c$                    & $J/\psi J/\psi$
            \\ \hline
            $cc\bar c\bar b$ & $\eta_cB_c^+$  & $J/\psi B_c^+, B_c^{*+}\eta_c$    & $J/\psi B_c^{*+}$
            \\ \hline
            $cc\bar b\bar b$ & $B_c^+B_c^+$   & $B_c^{*+}B_c^+$                   & $B_c^{*+}B_c^{*+}$
            \\ \hline
            $bb\bar c\bar b$ & $\eta_b B_c^-$ & $\Upsilon B_c^-, B_c^{*-}\eta_b$ & $\Upsilon B_c^{*-}$
            \\ \hline
            $bb\bar b\bar b$ & $\eta_b\eta_b$ & $\Upsilon \eta_b$                 & $\Upsilon\Upsilon$
            \\ \hline\hline
        \end{tabular}}
    \label{tab:coupled-channels}
\end{table}

\subsection{The $cc\bar{c}\bar{c}$ system}

In the $cc\bar{c}\bar{c}$ system, the $PP$ interaction involves only one channel $\eta_c\eta_c$, the $VP$ interaction involves only one channel $J/\psi\eta_c$, and the $VV$ interaction also involves only one channel $J/\psi J/\psi$.
 Besides, the $bb\bar{b}\bar{b}$ system can be similarly investigated.

Since the two vertices are both zero, $\mathcal{L}_{\eta_c \eta_c J/\psi} = 0$ and $\mathcal{L}_{J/\psi J/\psi J/\psi} = 0$, the interactions in this system all vanish within the extended local hidden gauge framework:
\begin{align}
\label{eq:coeff-cccc}
C_{PP}^t &= C_{PP}^u = 0 \, , \nonumber \\
C_{VP}^t &= C_{VP}^u = 0 \, , \nonumber \\
C_{VV}^t &= C_{VV}^u = 0 \, , \nonumber \\
C_{VV}^{co} &= 0 \, .
\end{align}

The $bb\bar{b}\bar{b}$  system and the  $cc\bar{c}\bar{c}$  system exhibit very similar dynamic properties, with the only distinction being their mass differences. Consequently, they share the same coefficients listed in Eq.~\ref{eq:coeff-cccc}, implying that the interactions within the  $bb\bar{b}\bar{b}$  system also vanish.

Therefore, our results do not support the existence of hadronic molecules in the $cc\bar{c}\bar{c}$ and $bb\bar{b}\bar{b}$ systems. Moreover, our conclusions actually do not depend on the value of the coupling constant $g$. As long as the near-threshold interactions in these systems are dominated by vector meson exchange and can be described by the Lagrangian in Eq.~\ref{eq-lagrangians}, the validity of the conclusions can be maintained. Let's examine the specific form of the $t$-channel exchange potential in Eq.~\ref{eq:vpp},
\begin{align}
    -iV^t = ig_1 (p_1+p_3)^\mu i g_2 (p_2 + p_4)^\nu \frac{i(-g_{\mu\nu}+ \frac{q_\mu q_\nu}{m_V^2})}{t-m_V^2} \, ,
\end{align}
where $p_1$ and $p_2$ are four-momenta of the incoming mesons; $p_3$ and $p_4$ are four-momenta of the outgoing mesons; $q$ is four-momenta of the exchanged meson and $g_1$ and $g_2$ are two coupling constants corresponding to top and bottom vertex in the fig.~\ref{fig-lagrangians} (a). Ignoring the three-momenta of the particles and approximating $t$ as 0, the following simplification can be obtained,
\begin{align}
    V^t \simeq {{g_1 g_2}\over{m_V^2}} (m_1+m_3)\cdot (m_2 + m_4).
\end{align}
It can be found that if the particles in the channels are identical, i.e., $g_1 = g_2$, the interaction will be repulsive. Similar derivations also apply to the exchange potential $V^u$ in the $u$-channel. This indicates that the existence of near-threshold molecular states in channels similar to $\eta_c \eta_c$ is difficult to support. The experimentally observed states in these systems are therefore good candidates to be tetraquark states.

\subsection{The $cc\bar{c}\bar{b}$ and $\bar{c}\bar{c}cb$ systems}

The results for the $cc\bar{c}\bar{b}$ and $\bar{c}\bar{c}cb$ systems are the same, so we only need to study the $cc\bar{c}\bar{b}$ system.  Besides, the $bb\bar{c}\bar{b}$ and $\bar{b}\bar{b}cb$ systems can be similarly investigated. In the $cc\bar{c}\bar{b}$ system, the $PP$ interaction involves only one channel $\eta_c B_c^+$, whose coefficients are
\begin{eqnarray}
    C_{PP}^t &=& 0 \, , \\
    \nonumber C_{PP}^u &=& \lambda\frac{1}{m^2_{B_c^{*}}} \, .
\end{eqnarray}
 The reduction factor $\lambda$ existing in the $u$ channel is introduced to account for the large mass difference between the initial meson $\eta_c$ and the final meson $B_c^+$ (or between the initial meson $B_c^+$ and the final meson $\eta_c$). Following Ref.~\cite{Yu:2018yxl}, numerically we use
\begin{equation}
    \lambda_{\eta_c B_c^+ \to B_c^+ \eta_c} \approx \frac{-m_{B_c^*}^2}{(m_{\eta_c}-m_{B_c})^2-m_{B_c^*}^2} = 1.37.
\end{equation}
The coefficient $C_{PP}^u$ is positive, indicating the interaction due to the exchange of the $B_c^{*}$ meson is repulsive, hence hadronic molecules in the PP sector are not expected to exist.

The $VP$ interaction involves two coupled channels $J/\psi B_c^+$ and $B_c^{*+} \eta_c$, whose coefficient are
\begin{eqnarray}
    \label{eq:cccbVP}
    \setlength{\arraycolsep}{5pt}
    \renewcommand{\arraystretch}{2}
    C_{VP}^t &=& \left(
    \begin{array}{c|cc}
            J=1             & J/\psi B_c^+                 & B_c^{*+} \eta_c
            \\ \hline
            J/\psi B_c^+    & 0                            & \lambda\frac{1}{m_{B_c^*}^2} \\
            B_c^{*+} \eta_c & \lambda\frac{1}{m_{B_c^*}^2} & 0
        \end{array}
    \right)\,\quad and
    \\
    \nonumber C_{VP}^u &=& {\bf 0}_{2 \times 2} \, .
\end{eqnarray}
Diagonalizing this $2 \times 2$ matrix, we obtain two decoupled channels:
\begin{eqnarray}
    |VP^+ \rangle &=& \frac{1}{\sqrt{2}}\left(|J/\psi B_c^+ \rangle + |B_c^{*+} \eta_c \rangle\right),
    \label{def:1bVP+}
    \\ |VP^- \rangle &=& \frac{1}{\sqrt{2}}\left(|J/\psi B_c^+ \rangle - |B_c^{*+} \eta_c \rangle\right),
\end{eqnarray}
whose coefficient is
\begin{eqnarray}
    \setlength{\arraycolsep}{5pt}
    \renewcommand{\arraystretch}{2}
    C_{VP}^{\prime t} &=& \left(
    \begin{array}{c|cc}
            J=1  & VP^+                         & VP^-
            \\ \hline
            VP^+ & \lambda\frac{1}{m_{B_c^*}^2} & 0                             \\
            VP^- & 0                            & -\lambda\frac{1}{m_{B_c^*}^2}
        \end{array}
    \right)\, .
\end{eqnarray}
Hence, the interaction due to the exchange of the $B_c^{*}$ meson in the $VP^-$ channel turns out to be  attractive, so there may exist a hadronic molecule of $J^P=1^+$ in the $VP$ sector.

The $VV$ interaction involves only one channel $J/\psi B_c^{*+}$, whose coefficients are
\begin{eqnarray}
    C_{VV}^t &=& 0 \,\quad and \\
    \nonumber C_{VV}^u &=& \lambda\frac{1}{m^2_{B_c^{*}}} \, .
\end{eqnarray}
The relevant contact term is
\begin{eqnarray}
    V^{co}_{J/\psi B_c^{*+} \to J/\psi B_c^{*+}}(s) &=&
    \left\{\begin{array}{cc}
        -2 g^2 & ~\textrm{for $J=0$}, \\
        3  g^2 & ~\textrm{for $J=1$}, \\
        g^2    & ~\textrm{for $J=2$}.
    \end{array}\right.
\end{eqnarray}
After performing the spin projection, we find the $J=1$ channel to be attractive, so there may exist a hadronic molecule of $J^P=1^+$ in the $VV$ sector.

\subsection{The $cc \bar{b} \bar{b}$ and $\bar c\bar c bb$ systems}

The results for the $cc \bar{b} \bar{b}$ and $\bar c\bar c bb$ systems are the same, so we only need to study the $cc \bar{b} \bar{b}$ system.  In this system, the $PP$ interaction involves only one channel $B_c^+ B_c^+$, whose coefficients are
\begin{eqnarray}
    C^t_{PP} &=& \frac{1}{m^2_{J/\psi}} + \frac{1}{m^2_{\Upsilon}} \, \quad and \\ \nonumber
    C^u_{PP} &=& \frac{1}{m^2_{J/\psi}} + \frac{1}{m^2_{\Upsilon}} \, .
\end{eqnarray}
The $VP$ interaction involves only one channel $B_c^{*+} B_c^+$, whose coefficient are
\begin{eqnarray}
    C^t_{VP} &=& \frac{1}{m^2_{J/\psi}} + \frac{1}{m^2_{\Upsilon}} \, \quad and
    \\
    \nonumber C_{VP}^u &=& 0 \, .
\end{eqnarray}
The $VV$ interaction also involves only one channel $B_c^{*+} B_c^{*+}$, whose coefficients are
\begin{eqnarray}
    C^t_{VV} &=& \frac{1}{m^2_{J/\psi}} + \frac{1}{m^2_{\Upsilon}} \, \quad and \\ \nonumber
    C^u_{VV} &=& \frac{1}{m^2_{J/\psi}} + \frac{1}{m^2_{\Upsilon}} \, .
\end{eqnarray}
The relevant contact term is
\begin{eqnarray}
    V^{co}_{B_c^{*+}B_c^{*+} \to B_c^{*+}B_c^{*+}}(s) &=&
    \left\{\begin{array}{cc}
        -8 g^2 & ~\textrm{for $J=0$}, \\
        0      & ~\textrm{for $J=1$}, \\
        4 g^2  & ~\textrm{for $J=2$}.
    \end{array}\right.
\end{eqnarray}
In sectors other than the $|(VV)_{cc\bar b\bar b}; J^{P}=0^{+}\rangle$  sector, the coefficients are positive, corresponding to repulsive interactions. As for the $|(VV)_{cc\bar b\bar b}; J^{P}=0^{+}\rangle$  sector, although the contact term provides an attractive potential of $-8g^2$, the repulsive potential generated by the exchange of vector mesons is approximately $+13g^2$ at threshold, resulting in an overall repulsive interaction. Therefore, in these sectors, the exchange of vector mesons cannot bind mesons together, hence the above coefficients do not support the existence of hadronic molecules in the $cc \bar{b} \bar{b}$ system.

\section{Numerical Results}
\label{sec:result}

In the previous section we have studied the interactions of the $cc\bar c\bar c$, $cc\bar c\bar b/\bar c\bar c cb$, $cc\bar b\bar b/\bar c\bar c bb$, $bb\bar c\bar b/\bar b\bar b cb$, and $bb\bar b\bar b$ systems. In this section we numerically study their properties. As shown in Eq.~(\ref{eq-cutoff-G}), the loop function $G(s)$ is regularized using the cutoff method, with the cutoff momentum $\Lambda$ describing the dynamical scale to be integrated out. Its value is quite uncertain for the exchange of fully-heavy vector mesons, and we follow Ref.~\cite{Liu:2023gla} to choose a broad region, $\Lambda=400\sim1400\mev$, to perform numerical analyses,
 since the authors of Refs.~\cite{Lu:2014ina,Ozpineci:2013zas} have already found that the requirement of heavy quark symmetry demands the use of the same cutoff momentum in the charm and bottom sectors. We note that the value of this important parameter is quite uncertain for the exchange of fully-heavy vector mesons, so the present study as well as Ref.~\cite{Liu:2023gla} serve as pioneer researches to investigate the fully-heavy hadronic molecules, but there do exist large theoretical uncertainties.

Within the extended local hidden gauge framework, the resonances are dynamically generated as poles of the scattering amplitudes $T_{PP/VP/VV}(s)$. We find the existence of eight poles that may lead to some singular structures on the invariant mass spectrum.
 There exist two poles in the $cc\bar c\bar b$ system and two poles in the $bb\bar c\bar b$ system, which we shall discuss in details later. We summarize their positions in Table~\ref{tab:pole} with respect to the cutoff momentum $\Lambda$. Besides, there exist four charge-conjugated poles in the $\bar c\bar c cb$ and $\bar b\bar b cb$ systems.

\begin{table*}[htb]
    \centering
    \caption{Pole positions with respect to the cutoff momentum $\Lambda$, in units of MeV. We only list the poles that correspond to the sub-threshold bound states.}
    \label{tab:pole}
    \setlength{\tabcolsep}{3.6pt}
    \renewcommand\arraystretch{1.7}
    \begin{tabular}{ccccccc}
        \hline
        \hline
        Pole                                            & $\Lambda = 400$     & $\Lambda = 600$    & $\Lambda = 800$   & $\Lambda = 1000$           & $\Lambda = 1200$       & $\Lambda = 1400$    \\
        \hline \hline
        $|(VP)_{cc\bar c\bar b}; J^{P}=1^{+}\rangle$        & $--$    & $--$               & $--$& $--$     & $--$ & $--$ \\
        \hline
        $|(VV)_{cc\bar c\bar b}; J^{P}=1^{+}\rangle$       & $--$                & $--$       & $--$ & $--$ & $--$ & $--$ \\
        \hline
        $|(VP)_{bb\bar c\bar b}; J^{P}=1^{+}\rangle$       & $--$     & $--$   & $--$ & $15725.3-i0$    & $15710.8-i0$     & $15685.2-i0$ \\
        \hline
        $|(VV)_{bb\bar c\bar b}; J^{P}=1^{+}\rangle$      & $--$                & $--$               & $--$ & $--$ & $15790.3-i0$ & $15784.0-i0$ \\
        \hline
        \hline
    \end{tabular}
\end{table*}

We find two poles in the $cc\bar c\bar b$ system: one pole in the $VP$ sector and the other in the $VV$ sector. However, both of them correspond to virtual states when setting the cutoff momentum $\Lambda=400\sim1400\mev$, so they can only result in some threshold effects. The pole in the $VP$ sector corresponds to the sub-threshold bound state with $\Lambda > 1550\mev$, and the pole in the $VV$ sector corresponds to the sub-threshold bound state with $\Lambda > 2650\mev$. We generally consider the cutoff momentum $\Lambda$ to be consistent with the chiral unitary approach, which takes $\Lambda \approx 4 \pi f_\pi \simeq 1200$ MeV. This value reflects certain non-perturbative properties of QCD. However, it is important to note that the parameter $\Lambda$ also functions as a free parameter, absorbing some implicitly considered interactions, and as a result, it may deviate from 1200 MeV in practical applications. Empirically, $\Lambda$ is usually taken to fall within the range of $400$ to $700$ MeV. Given the uncertainties inherent in our work, we believe it is appropriate to extend this range to $400$ to $1400$ MeV. Therefore, our results do not support the existence of deeply-bound hadronic molecules in the $cc\bar c\bar b$ system.

\begin{figure*}[htbp]
    \centering
    \subfloat[$T_{VP}^2$ with $\Lambda = 750\mev$]{\includegraphics[width=.32\linewidth]{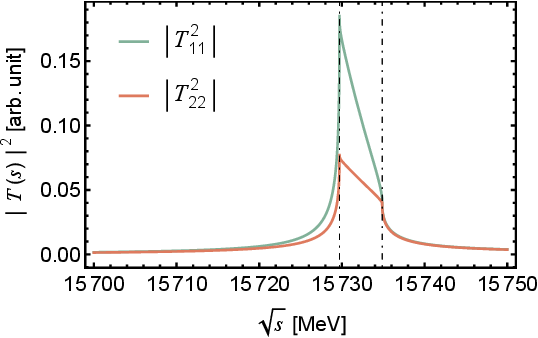}}\hspace{5pt}
    \subfloat[$T_{VP}^2$ with $\Lambda = 850\mev$]{\includegraphics[width=.3\linewidth]{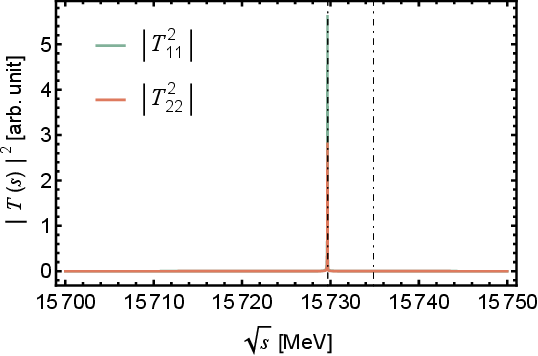}}\hspace{5pt}
    \subfloat[$T_{VP}^2$ with $\Lambda = 950\mev$]{\includegraphics[width=.31\linewidth]{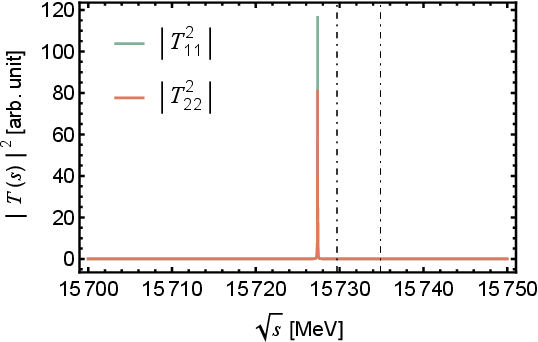}}\\
    \subfloat[$T_{VV}^2$ with $\Lambda = 1000\mev$]{\includegraphics[width=.3\linewidth]{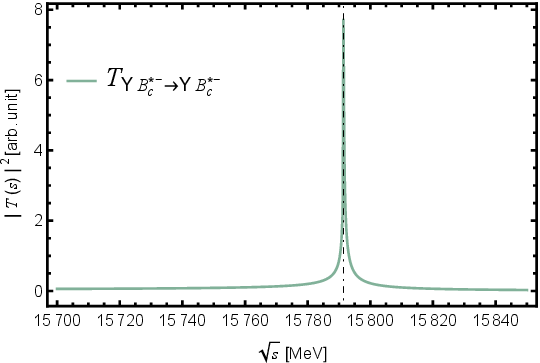}}\hspace{5pt}
    \subfloat[$T_{VV}^2$ with $\Lambda = 1100\mev$]{\includegraphics[width=.305\linewidth]{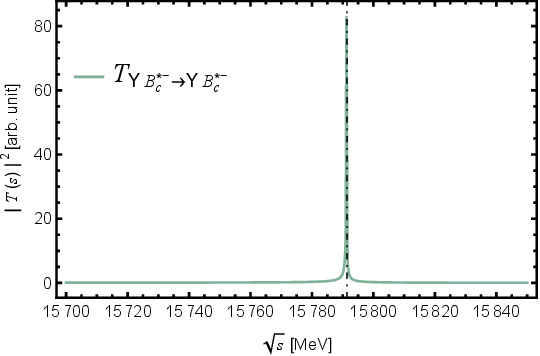}}\hspace{5pt}
    \subfloat[$T_{VV}^2$ with $\Lambda = 1200\mev$]{\includegraphics[width=.315\linewidth]{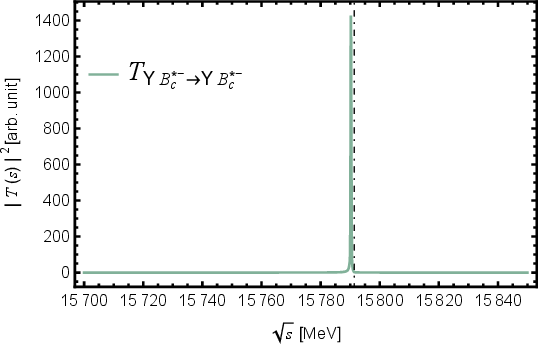}}
    \caption{Line shapes of the transition amplitudes $\lvert T(s) \rvert^2$ for the cutoff momentum $\Lambda = $ (a)~750\mev, (b)~850\mev, and (c)~950\mev\  in the $VP$ sector as well as $\Lambda = $ (d)~1000\mev, (e)~1100\mev, and (f)~1200\mev\  in the $VV|_{J=1}$ sector of the $bb\bar c\bar b$ system. The relevant thresholds are indicated by dashed lines. In the subfigures~(a,b,c) the green line labeled as $T_{11}$ and the red line labeled as $T_{22}$  represent $|T_{\Upsilon B_c^- \to \Upsilon B_c^-}(s)|^2$ and $|T_{B_c^{*-}\eta_b \to B_c^{*-}\eta_b}(s)|^2$, respectively.}
    \label{fig-result-01}
\end{figure*}

We also find two poles in the $bb\bar c\bar b$ system: one pole in the $VP$ sector and the other in the $VV$ sector. The pole in the $VP$ sector corresponds to the sub-threshold bound state when setting $\Lambda > 850\mev$, making it possible to be identified as a hadronic molecule. This pole transfers to a virtual state  and results in the threshold effect when setting $\Lambda < 850\mev$. The pole in the $VV$ sector corresponds to the sub-threshold bound state with $\Lambda > 1100\mev$. To illustrate these two poles, we present in Fig.~\ref{fig-result-01} the transition amplitudes using several different values of the cutoff momentum $\Lambda$.

\section{Conclusion}
\label{sec:con}

In this paper we study the fully-heavy meson-meson interactions with the quark constituents $cc\bar c\bar c$, $cc\bar c\bar b/\bar c\bar c cb$, $cc\bar b\bar b/\bar c\bar c bb$, $bb\bar c\bar b/\bar b\bar b cb$, $bb\bar b\bar b$ through the extended local hidden gauge formalism. After solving the coupled-channel Bethe-Salpeter equation, we search for poles on both the first (physical) and second Riemann sheets. The obtained results are summarized in Table~\ref{tab:pole} with respect to the cutoff momentum $\Lambda$.
 We find two poles in the $bb\bar c\bar b$ system (and two charge-conjugated poles in the $\bar b\bar b cb$ system): the pole generated in the $VP$ sector corresponds to the sub-threshold bound state when setting the cutoff momentum $\Lambda > 850 \mev$, and the pole generated in the $VV$ sector corresponds to the sub-threshold bound state with $\Lambda > 1100 \mev$. These two poles are potential fully-heavy hadronic molecules, and we propose to investigate them in the $\mu^+\mu^-B_c^-$ channel at LHC. However, our results do not support the existence of hadronic molecules in the $cc\bar c\bar c$, $cc\bar c\bar b/\bar c\bar c cb$, $cc\bar b\bar b/\bar c\bar c bb$, and $bb\bar b\bar b$ systems.

Besides, the $c b \bar c \bar b$ system has already been investigated in our previous study~\cite{Liu:2023gla}, where we found the existence of the fully-heavy hadronic molecules $|B_c^{+} B_c^{-}; J^{PC}=0^{++} \rangle$, $|B_c^{*+} B_c^{-} - c.c.; J^{PC}=1^{+-} \rangle$, and $|B_c^{*+} B_c^{*-}; J^{PC}=2^{++} \rangle$ as well as the possible existence of $|B_c^{*+} B_c^{-} + c.c.; J^{PC}=1^{++} \rangle$. We further proposed in Ref.~\cite{Liu:2023gla} that a lower-mass fully-heavy meson may be able to bind two higher-mass fully-heavy hadrons. The results obtained in the present study are consistent with this proposal: the exchanged mesons of the $bb\bar c\bar b$ and $cc\bar c \bar b$ systems are both the $B_c^*$ meson, but the larger mass of the $bb\bar c\bar b$ system facilitates the formation of bound states. It is a topic of considerable interest whether the interaction of the heavy meson exchange is strong enough to form hadronic molecules. This question serves as a crucial test for the extensively investigated interaction of the light meson exchange. Therefore, the present study as well as Ref.~\cite{Liu:2023gla}, both of which concentrate on the interaction of the fully-heavy meson exchange, are of particular interest.

\section*{Acknowledgments}

This project is supported by
the National Natural Science Foundation of China under Grant No.~12075019,
the Jiangsu Provincial Double-Innovation Program under Grant No.~JSSCRC2021488,
and
the Fundamental Research Funds for the Central Universities.

\bibliographystyle{elsarticle-num}
\bibliography{ref}

\end{document}